\documentstyle[11pt]{article}

\input{mssymb}
\newcommand{\got}{\frak}
\newcommand{\emp}{\emptyset}
\newtheorem{thm}{Theorem}[section]
\newtheorem{prop}[thm]{Proposition}

\begin{document}

{\huge Residues in nonabelian localization}

\begin{center}
     Jaap Kalkman \footnote{supported by a
     NATO-Science Fellowship}\\
     MIT\\
     Department of Mathematics\\
     Cambridge MA 02139
\end{center}
\begin{center} \today \end{center}

\section{introduction}
In this paper we would like to elaborate on the formula, recently
obtained by Jeffrey and Kirwan [JK], that expresses integrals over the reduced
phase space as a sum of residues over fixed points. This elaboration might be
useful in order to obtain information on
the cohomology ring of the reduced phase
space.

Let $G$ be a compact Lie group acting Hamiltonially on a
compact symplectic manifold $M$ with
momentum map $\mu:M \rightarrow {\got g}^*$. Let us assume that $0$ is a
regular
value for this map and denote the reduced phase space by $M_0=\mu^{-1}(0)/G$.
$M_0$ is
an orbifold. Furthermore, fix a maximal torus $T \subset G$ and
assume that the fixed points $P_j$ for the $T$ action are isolated.
The weights $\lambda^i(P_j)$ of the torus action at the
fixed points and the values $\mu(P_j)$ of the momentum mapping of the $T$
action
each determine a hyperplane in ${\got t}$.
Let $Y \in {\got t}$ be a fixed vector that does not lie in any of
these hyperplanes.

Let $F_Y=\{ P_j \mid (\mu(P_j),Y)<0\}$. For any equivariantly closed form
$\alpha$
of degree dim($M_0$) on $M$
Jeffrey and Kirwan proved the following formula:
\begin{equation}
  \int_{M_0} r (\alpha) = \sum_{P_j \in F_Y} {\rm Res} (e^{i\;\mu(P_j)(X)}
    \left( \frac{\pi(X)
\iota_{P_j}^*\alpha(X)}{\prod_{i=1}^{n}\lambda^i(P_j)(X)}
    \right) )
\end{equation}
where $r$ is the natural projection map to differential forms on $M_0$, $\pi$
is the product
of roots associated to $T \subset G$ and Res is a residue, introduced by
Jeffrey
and Kirwan in [JK]. This residue depends on $Y$.

It is these residues that we would like to understand better. We will see
that they can be expressed in terms of determinants of the weights
$\lambda^i(P_j)$
and the values
$\mu(P_j)$. Furthermore,
we will generalize (1) to non-compact manifolds using [PW]
and apply this to torus actions on ${\bf C}^n$.

\section{preliminaries}

Let $M$ be a manifold on which a compact Lie group $G$ acts.
Let $X_a$ be a basis for the Lie algebra ${\got g}$ and let
$\phi^a$ be the dual basis, regarded as generators of
$S({\got g}^*)$, the polynomials on ${\got g}$. The equivariant
cohomology associated to the $G$-action on $M$ can be modeled
using the Cartan model, which we will describe briefly now.
Note that this model only works for compact $G$ ([AB]).

The algebra of the Cartan model is
\[ (S({\got g}^*) \otimes
   \Omega (M))^G.
\]
On $\Omega (M)$, the differential forms on $M$,
there is a super Lie algebra of derivations associated to the
$G$-action on $M$. It is spanned by d, the exterior differential
of degree 1, by $\iota_a$, the inner products of degree -1 on
forms with vector fields generated by the $X_a$, and by ${\cal L}_a$,
the Lie derivatives of degree zero on forms using the same
vector fields (see [Ka] for more details).

The differential on this Cartan algebra is
\[ {\rm D}=1 \otimes {\rm d} -
   \phi^a \otimes \iota_a,
\]
where a summation over the index $a$ is
understood. This differential squares to zero and the associated
cohomology is called the equivariant cohomology $H_G(M)$. It is
graded if we declare deg($\phi^a$)=2, so that deg(D)=1.

Equivariant cohomology has a lot of interesting properties. The
extremes are, for a (locally) free action
\begin{equation}\label{freeisom}
 H_G(M) \simeq H(M/G)
\end{equation}
and for a
trivial action
\begin{equation}\label{trivisom}
 H_G(M) \simeq S({\got g}^*)^G \otimes H(M).
\end{equation}

Hamiltonian group actions are closer to being trivial than being
free, since in this case $H_G(M)$ inherits a filtration such that
the associated graded algebra equals $S({\got g}^*)^G \otimes
H(M)$. ([Ki], [Gi]).

Furthermore, there are a couple of interesting formulas using
equivariant cohomology, the most important being the
localization formula ([AB],[BV]).

Let $T \subset G$ be a maximal torus and ${\cal F} \subset M$ its
fixed point set. If $\alpha \in (S({\got g}^*) \otimes \Omega(M))^G$ is
equivariantly closed, then
\begin{equation}\label{ABBV}
     \int_M \alpha = \sum_{F \subset {\cal F}} \int_F
     \frac{\iota^*_F \alpha}{\epsilon_F}
\end{equation}
where $\int_M: (S({\got g}^*) \otimes \Omega(M))^G \rightarrow
 S({\got g}^*)^G$ extends integration over $M$, $\epsilon_F$
 represents the equivariant Euler class of $F \subset M$, and
 the summation is over the connected components of ${\cal F}$.
 If there are no fixed points, integration over $M$ will always give
 zero.

One can extend this localization formula to manifolds with
boundary. Let $K \subset G$ be a circle subgroup  in the
center of $G$ that preserves
the boundary and, moreover, acts locally free on the
boundary $\partial M$.
We get the following extension:
\begin{equation}\label{locbound}
     \int_M \alpha = \sum_{F \subset {\cal F}} \int_F
     \frac{\iota^*_F \alpha}{\epsilon_F}
     -\int_{\partial M} \frac{\theta \wedge \alpha}{{\rm D}\theta}
\end{equation}
where $\theta$ is a connection form for the $K$ action on the
boundary.

Let $r: H_G(M) \rightarrow H_{G/K}(\partial M/K)$
be restriction to $\partial M$ followed by (\ref{freeisom}).
Alternatively, this map can be described as restriction of
$\alpha$ to $\partial M$ and then taking the residue of
$\frac{\theta \wedge \alpha}{{\rm D} \theta}$. Taking the residue
is by definition taking the coefficient of $\frac{1}{x}$, where
$x$ is the variable in $S({\got g}^*)$ corresponding to the
circle $K$. Of course this cannot be done without choosing a
splitting of ${\got g}$. However, due to the topological
character of the residue, the result will be independent of
such a splitting.

Taking the residue everywhere in (\ref{locbound}) gives us
\begin{equation}\label{jkcircle}
   \int_{\partial M/K} r(\alpha) = \sum_{F \subset {\cal F}}
       \int_F {\rm res} \frac{\iota_F^* \alpha}{\epsilon_F}.
\end{equation}
The lhs of (\ref{locbound}) vanishes since it is a polynomial
rather than a fractional expression.

For $G=K$ this formula was proved in [Ka]. A special case of this
would be if the circle $G$ acts Hamiltonially with momentum map
$\mu$ and $M=\{x \mid \mu(x) \geq 0\}$. Then $\partial M/K$ is the
reduced phase space and we obtain an expression for integrals over
the reduced phase space in terms of certain fixed points on the
original manifold ([Wu]).

An idea of how to do this in general for Hamiltonian non-abelian compact
group actions is due to Witten ([Wi]) and worked out by Jeffrey
and Kirwan in [JK]. They obtained the formula stated in the introduction.
In fact, their formula is more general since it does not require
the fixed points to be isolated.

By definition their residue is an integration over ${\got t}+iY$.
To understand it, we could not do better than compute it by hand,
in the case of isolated fixed points. Also, we will assume some
genericity conditions. In general, the residue might depend on
the choice of a test function, although the sum, of course, does
not (see, e.g., [Du]).

\section{Geometry at an isolated fixed point}

Let $P$ be an isolated fixed point of the $T$ action on $M$.
Let $r={\rm dim}(T)$ and $2n={\rm dim}(M)$. The infinitesimal
action of ${\got t}$ on $T_PM$ gives rise to $n$ weights $\lambda^1, \ldots,
\lambda^n \in {\got t}^*$ together with their hyperplanes $V_i$ in ${\got t}$.
We will assume that these weights satisfy the folowing genericity condition:
all $r$-tuples are independent.

Let us denote $\mu(P)$ simply by $\mu \in {\got t}^*$.
Choose a basis $v^1, \ldots, v^r$ for ${\got t}^*$ with $v^r=\mu$ and
such that these vectors, together
with the weights, still satisfy the genericity condition. In particular,
this implies that for any $(r-1)$ tuple of weights ($\lambda^{j_i}$),
det$(\lambda^{j_1}, \ldots, \lambda^{j_{r-1}}, \mu) \neq 0$.

We will use this basis to successively perform the $r$ integrations and finally
end up with an explicit expression for the residue. Associated to this basis
there is a dual basis $v_1, \ldots, v_r$ for ${\got t}$ defined through
$(v^i,v_j)=\delta_j^i$.

In the sequel, we will need the halfspaces in ${\got t}$
\begin{equation}\label{hsp}
	   h_j : \hspace{20pt} <v_1, \ldots, v_{j-1}> +
                {\bf R}_{\geq 0}v_j + Y, \hspace{20pt} j=1, \ldots, r-1,
\end{equation}
lying in the hyperplane $
V_\mu = \{ X \in {\got t}  \mid (\mu,X)=(\mu,Y) \}$ through the
fixed vector $Y \in {\got t}$. Note that dim$(h_j)=j$ and that this is the
standard half-flag given by the basis, shifted by $Y$.

\vspace{10pt}

{\bf Definition.} An ordered $(r-1)$-tuple $V_{j_1}, \ldots, V_{j_{r-1}}$ of
weight hyperplanes is called {\it positive} with respect to the shifted
half-flag $(h_1, \ldots ,h_{r-1})$ if the following chain of conditions holds:
\begin{eqnarray}
  V_{j_1} \cap h_1 \neq \emp \hspace{20pt} (C1) \nonumber \\
  (V_{j_1} \cap V_{j_2}) \cap h_2 \neq \emp \hspace{20pt} (C2) \nonumber \\
  \vdots  \hspace{25pt} \vdots \hspace{5pt} \nonumber \\
  (V_{j_1} \cap \ldots \cap V_{j_{r-1}}) \cap h_{r-1} \neq \emp
	 \hspace{20pt} (Cr) \nonumber
\end{eqnarray}

Note that, because of the genericity condition, the intersections above
are either empty or contain a single point not lying on the
boundary of the halfspace
(by slightly moving $Y$, if necessary).
In particular, this point is always different
from $Y$, which also follows from the fact that $Y$ is not
contained in any of the weight hyperplanes.

In the next section we will see that the residue  at $P$ is a sum over $(r-1)$
tuples $I$ of weights, and, moreover, that only positive tuples contribute.
Therefore, it would be most convenient to use a basis of ${\got t}$ that
leads to the smallest number $N$ of tuples that have positive orderings.
We will find a general upperbound for $N$, smaller than
$\left( \begin{array}{c} n \\ r-1 \end{array} \right) $, in section six.

To finish this section we introduce some notations useful in the sequel.
Let $I$ be a (non-ordered) set $\{j_1, \ldots, j_{r-1} \} \subset
\{ 1, \ldots, n\}$. We denote by $n_I$ the number of positive $(r-1)$
tuples of weight hyperplanes having the same index set $I$.
This number is useful because it turns out that any positive tuple
wih the same index set $I$ contributes to the residue in the same way.
We also associate to each set $I$ the element $Q_I=\sum q^i v_i$ in ${\got t}$,
with coordinates
\begin{equation}\label{kui}
  q^i = \frac{{\rm det}(\lambda^{j_1}, \ldots, \lambda^{j_{r-1}}, v^i)}
	     {{\rm det}(\lambda^{j_1}, \ldots, \lambda^{j_{r-1}}, \mu)}.
\end{equation}
We will use that $\lambda^j(Q_I)\neq 0$ as long as $j\notin I$. Again, this
follows from the assumption that we called genericity condition.
Note that $Q_I$ does not depend on our choice of a basis for ${\got t}$.

\section{An expression for the residue}

Using the notation and assumptions of the previous sections, we will
prove the following version of the nonabelian fixed point formula.

\begin{thm}
	Let $M$  be a compact symplectic manifold
	with a Hamiltonian action of a compact Lie group $G$.
	Assume that $0 \in {\got g}^*$ is a regular value of the
	momentum mapping so that the reduced phase space $M_0$ exists
	as an orbifold. Let $T$ be a maximal torus inside $G$ and
	assume that the fixed points $P_i$ of the $T$-action are
	all isolated and have weights $\lambda^j(P_i)$. Then,
	for any closed equivariant differential form $\alpha$ on $M$
	of degree dim$(M_0)$, we have
\begin{equation}\label{jkexpl}
  \int_{M_0} r (\alpha) = c \sum_{P_i \in F_Y} \sum_I
   n_I \frac{\pi(Q_I)\alpha(Q_I)}{\prod_{j \notin I} (\lambda^j(P_i),Q_I)}.
\end{equation}
where the second summation is over all $(r-1)$ tuples of weights and
	$n_I$ and $Q_I$ are as defined above.
\end{thm}

{\bf Proof.}
For each fixed point $P_j$ we have to compute
\begin{equation}
  \int_{{\got t}+iY} e^{i(\mu,X)} \frac{\pi(X)\alpha(X)}
					{\prod_{l=1}^n (\lambda^l,X)}
					{\rm d}X.
\end{equation}
Using the coordinates of the previous section, we have $(\mu,X)=X^r$.
We will perform the integration by successive integration over
$X^j, \; j=1, \ldots ,r$. Each integration is over a line
${\rm Im}(z^j)=Y^j$ in ${\got t}_{\bf C}$, which we will close up
to  a cycle in order to be able to apply Cauchy's integral formula.
Here, $z^j$ is the coordinate corresponding to the complexification
of the $X^j$ axis.
For $j=1, \ldots, r-1$, we have the choice of closing up in the upper
or lower halfplane. This is because the integral over the added part
vanishes,
using, e.g., an expanding box with basis ${\rm Im}(z^j)=Y^j, \; \mid X^j \mid
\leq R, \; R \rightarrow \infty$.

For the very last integration
\begin{equation}
	\int_{{\bf R}+iY^r} \frac{e^{iX^r}}{X^r}
\end{equation}
we have to close the integration line into the upper half plane for
obvious reasons. Furthermore, from this one sees that if
$Y^r=(\mu,Y)>0$, then this integral vanishes, because the pole does
not lie inside the integration area. Hence the first summation,
only over the fixed points in $F_Y=\{P_j \mid (\mu(P_j),Y)<0\}$.

Let us close the first line in the upper half plane. All linear
equations $(\lambda^j,X)=0$, $(X^2, \ldots ,X^r $ are fixed) in
$X^1 \in {\bf C}$ give rise to poles. These poles lie inside the integration
area iff condition $(C1)$ holds. So we get a sum over at most $n$ poles of
integrals with one variable less. Evaluating the integral over $X^2$ in
the same way one sees that the poles lie inside the integration area
iff condition $(C2)$ holds, and so on. This way, we would have ended up with
a lot of summands, were it not that some of them result in the same answer.
In order to see what answer one obtains for a particular sequence of
linear equations, note that we first express $X^1$ in terms of $X^2, X^3,
\ldots$ (using the first equation), then $X^2$ in terms of $X^3, X^4,
\ldots$ (using the second equation, but with $X^1$ eliminated), etc.
Actually, what we are doing this way is equivalent to solving
a system of $r-1$ inhomogeneous linear equations
$(\lambda^j,X)=0$, where $X^r$ should not be considered as a variable.
Using Cramers' rule, one can figure out easily that the poles are
finally given by the $q^i$ defined by (\ref{kui}). We see from
these considerations that the order of the linear equations is not of any
importance, hence the answer (\ref{jkexpl}).

\section{Torus actions on ${\bf C}^n$}

We would like to apply the previous result to torus actions on ${\bf C}^n$.
However, this space is not compact and therefore we cannot use our result
right away. We will make use of [PW] to overcome this.
In their paper, Prato and Wu showed that localization on non compact
spaces of specific equivariant differential forms, like the ones used in
[JK], still exists, provided one is willing to choose $Y$ more carefully.
More precisely, one has to choose $Y$ inside the dual of the asymptotic cone
of the momentum map. Making this more careful choice, a direct consequence
of [PW] is that (\ref{jkexpl}) is also valid for noncompact symplectic spaces
with proper momentum maps. In fact, for generic torus actions on ${\bf C}^n$,
zero will be the only fixed point and the condition on $Y$
makes sure that $0 \in P_Y$.

So in this particular case there is only one fixed point to sum over and
this term had better be complicated because it contains all information
to obtain the cohomology of (generic) toric varieties. We will repeat the
result of the previous section here for this special case.
Assume that the linear $T^r$ action on ${\bf C}^n$ is given by weights
$\lambda^1, \ldots , \lambda^n \in {\got t}^*$.
We can write $H_T({\bf C}^n) \simeq {\bf C}[\phi^1, \ldots ,\phi^r]$,
using the same basis for ${\got t}^*$ as in the previous section. Let
$\alpha=\phi^J$, where $J$ is a multi-index of length $n-r$.

\begin{thm}
\begin{equation}\label{jkspec}
\int_{M_0} r(\alpha) = \sum_I \frac{n_I \; q^J}
	 {\prod_{j \notin I} \lambda^j(Q_I)}
\end{equation}
where $n_I$ is the number of positive $r-1$ tuples of weight hyperplanes
having the index set $I$.
\end{thm}

To show how this formula works out, we will do an example for $r=2$.
Let $T=S^1 \times S^1$, the first circle acting on ${\bf C}^n$ with
exponents $(1, \ldots , 1)$; the second one with $(m_1, \ldots, m_n)$.
The weights $\lambda^i$ are $(1, m_i)$ and for $I=\{ i \}$ we have
$Q_i=(-m_i,1)$. Hence $\lambda^j(Q_i)=m_j-m_i$. Choose $Y=(0,1)$. $Y$
lies inside the dual cone of $\mu(0)=(0,1)$ and does not lie on the
weighthyperplanes as long as $m_i \neq 0$. For $\alpha = \tau^k
\phi^{n-k-2} \in {\bf C}[\tau,\phi]$ we obtain:

\begin{equation}\label{circleform}
 \int_{M_0} r(\tau^k \phi^{n-k-2}) = \sum_{i, m_i>0}
     \frac{(-m_i)^k}{\prod_{j \neq i} (m_j-m_i)}
\end{equation}
which is the same formula we would have obtained using the formula for
a circle action on ${\bf CP}^{n-1}$ and $\tau$ representing the
equivariant symplectic form in $H_{S^1}({\bf CP}^{n-1})$.

\section{Use of the formula}

To determine $H(M_0)$, we need to know the kernel of
the ringhomomorphism
$r:H_T(M) \rightarrow H(M_0)$. We will show how formula (\ref{jkspec})
can help us. To begin with let us state an easy consequence of
the surjectivity of the map $r$ ([Ki]) and
Poincar\'e duality on the orbifold $M_0$ ([Sa]).
\begin{prop}
$\alpha \in H_T(M)$ is in ${\rm ker}(r)$ iff $\int_{M_0} r(\alpha)
       r(\beta) = 0$ for all $\beta \in H_T(M)$.
\end{prop}
The idea is now to use (\ref{jkspec}) to determine when the integral
over $M_0$ vanishes. Let $\epsilon : H_T(M) \rightarrow {\bf C}^N $
be the map $
\alpha \mapsto (\alpha(Q_{I_1}), \ldots , \alpha(Q_{I_N}))$
where $N$ is the number of unordered $r-1$ tuples $I$ of weights,
such that $n_I \neq 0$. Obviously, ker$(\epsilon) \subset {\rm ker}(r)$.
Therefore, we would like to know the smallest $N$ possible. Going back
to the proof in section four, one sees that the signs of the basis vectors
can be arranged such that a given weight $\lambda$ never occurs in
a positive tuple. Hence $\left( \begin{array}{c} n-1 \\ r-1 \end{array}
\right) $ is an upperbound for the minimum of $N$. How many weights
$\lambda$ are there at most in general that are never included in
positive tuples? Again, going back to the signs of the basis vectors one
sees that this number equals $\frac{n}{2^{r-1}}$, rounded above.
Denoting this number by $k$ we get $\left( \begin{array}{c} n-k \\ r-1
\end{array} \right) $ as an upperbound for the minimal $N$. In the rank
two case, e.g., $N$ is at most $\frac{1}{2} n$, which one clearly
observes in (\ref{circleform}).

For a given $I$, it is easy to find $\alpha_I \in H^2_T(M)$ ($\alpha_I$
is a homogeneous polynomial of degree $1$ in $r$ variables), such that
$\alpha_I (Q_I) = 0$. Therefore,
\begin{equation}
    \alpha = \prod_{i=1}^N \alpha_{I_i} \; \in \; {\rm ker}(r)
\end{equation}
This will often be a trivial statement, since, in general,
${\rm deg}(\alpha)=N $ will be bigger than
$\frac{1}{2}{\rm dim}(M_0)=n-r$. However, for rank two
($r=2$) groups we get an interesting element. Note that we get
more elements like $\alpha$ if we choose another basis for
${\got t}$, inducing another definition of a positive tuple,
and hence another collection $\{I_i\}$.
The two elements $\alpha$ that we get this way in the rank two
case generate ker$(r)$ ([Ka]).
For higher rank tori it seems hard to get general results this way.

\vspace{30pt}

I would like to thank the mathematics department of MIT, and
especially Victor Guillemin for the hospitality and the symplectic
geometry group as a whole for helpful discussions. Furthermore,
I am grateful to Hans Duistermaat for asking me the question what the residue
at zero for a generic torus action on ${\bf C}^n$ looks like.

\section*{References}

\begin{description}
\item[[AB]] M. Atiyah, R. Bott, The moment map and equivariant
                 cohomology,
                 Topology 23 (1984) 1-28
\item[[BV]] N. Berline, M. Vergne, Classes carct\'eristiques
                \'equivariantes,
		C.R. Acad. Sci. Paris 295 (1982) 539-541
\item[[Du]] J.J. Duistermaat, Equivariant cohomology and stationary phase,
		  preprint 817, University of Utrecht, August 1993
\item[[Gi]] V.A. Ginzburg, Equivariant cohomologies and
                 K\"ahler geometry,
                 Funct. Anal. and its Appl. 21 (1987) 271-283
\item[[JK]] L.C. Jeffrey, F.C. Kirwan, Localization for non-abelian
                 group actions, alg-geom/9307001
\item[[Ka]] J. Kalkman, Cohomology rings of symplectic quotients,
		to appear in J. Reine Angew. Math.
\item[[Ki]] F.C. Kirwan, Cohomology of Quotients in Symplectic
                   and Algebraic Geometry, Math. Notes Vol. 31,
                   Princeton University Press 1984
\item[[PW]] E. Prato, S. Wu, Duistermaat-Heckman measures in a
		non-compact setting, alg-geom/9307005
\item[[Sa]] I. Satake, On the generalization of the notion of
                 manifold, Proc. Nat. Acad. Sci. USA 42 (1956) 359-363
\item[[Wi]] E. Witten, Two dimensional gauge theories revisited,
                 J. Geom. Phys. 9 (1992) 303-368
\item[[Wu]] S. Wu, An integration formula for the square of moment
                 maps of circle actions, hep-th/9212071

\end{description}

\end{document}